# Simulation of photodetection using finite-difference time-domain method with application to near-field subwavelength imaging based on nanoscale semiconductor photodetector array

Ki Young Kim, Boyang Liu, Yingyan Huang, and Seng-Tiong Ho

*Department of Electrical Engineering and Computer Science, Northwestern University,*

*Evanston, IL 60208, USA*

**Abstract** Simulation of detecting photoelectrons using multi-level multi-electron (MLME) finite-difference time-domain (FDTD) method with an application to near-field subwavelength imaging based on semiconductor nanophotodetector (NPD) array is reported. The photocurrents from the photodiode pixels are obtained to explore the resolution of this novel NPD device for subwavelength imaging. One limiting factor of the NPD device is the optical power coupling between adjacent detector pixels. We investigate such power coupling in the presence of absorbing media as well as the spatial distributions of the electric field and photoelectron density using the MLME FDTD simulation. Our results show that the detection resolution is about one tenth of the operating wavelength, which is comparable to that of a near-field scanning optical microscope based on metal clad tapered fiber.

**Key words:** FDTD simulation, nanoscale photodetector (NPD) array, photocurrent, subwavelength resolution

## 1. Introduction

Semiconductor photodetector array has various applications in industry and academics, such as product inspection, medical imaging, security screening, and analytical characterization and imaging. Recently, semiconductor photodetector array with nanometer scale pixels has been drawing more and more attentions due to its high detection resolution, fast response and easy integratability (Liu *et al.* 2007 and Kolb *et al.* 1995). To model such semiconductor photodetector array with nanometer scalesize and complex optical geometry, a more sophisticated modeling technique is required. This model should be able to account for the electromagnetic wave propagation within the detector, its interaction with the absorbing semiconductor medium, and the generation of photocurrent.

In this paper, we report a new method to simulate photodetection in semiconductor material using multi-level multi-electron finite-difference time-domain algorithm (Huang 2002; Huang and Ho 2006). In MLME-FDTD method, where Pauli exclusion principle and Femi-Dirac thermalization are incorporated into the rate equation for the semiconductor material system, multiple energy levels are used to describe the essential characteristics of the semiconductor band structures, which allow us to model the full semiconductor carrier dynamics with reasonable accuracy for typical applications. As a result, photocurrents generated by active semiconductor materials, one of the figures of merit for photodetectors, could be calculated and evaluated. Using MLME model, we investigate both the light propagation and the physical mechanisms of the photodetection via the semiconductor materials.

## 2. Simulation of photodetection using FDTD method

Conceptually, a photodetector can be modeled as a medium with two energy levels in which the photocurrent can be calculated from the rate of excitation of ground-state electrons from the ground level (level $|1\rangle$) to the upper level (level $|2\rangle$), which are subsequently returned back to the ground level through an external electric circuit as shown in Fig. 1.

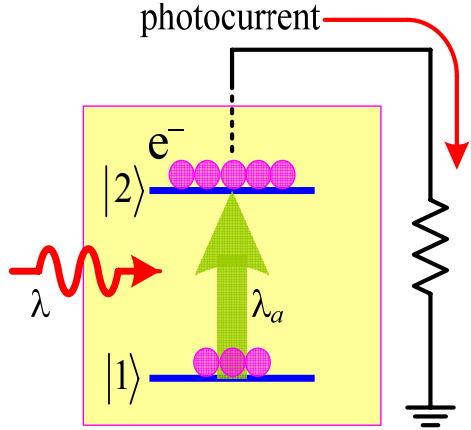

Fig. 1. Generation of photocurrent from photoelectrons in semiconductor. $\lambda = \lambda_a = 1550\text{nm}$ is assumed in this study. $\lambda$ and $\lambda_a$ are the incident wavelength and resonant wavelength of the semiconductor material, respectively.

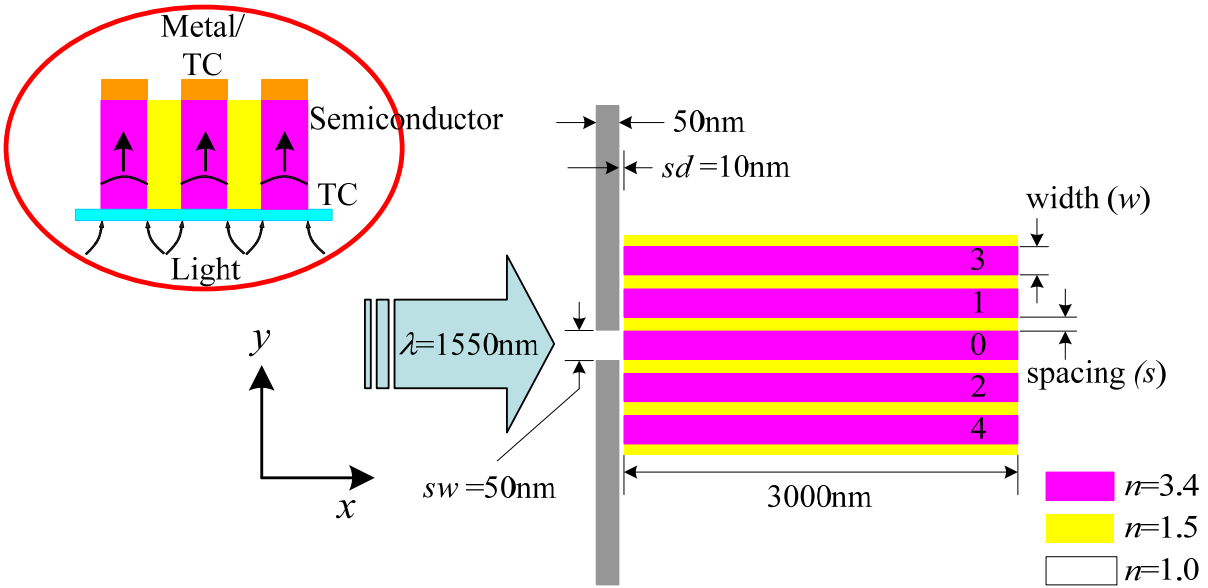

Fig. 2. Schematic of the NPD array. Light to be detected is from the subwavelength metal slit. The refractive indexes of the semiconductor, the filling dielectric material (benzocyclobutene, BCB), and air at 1550nm are assumed to be 3.4, 1.5, and 1.0, respectively. NPD pixels are labeled as 0, 1, 2, 3 and 4.

Fig. 2 shows a two-dimensional schematic of the NPD array for the FDTD simulation with its dimensions and operating parameters. Inset shows a practical photocurrent pickup mechanism. Transparent conductor (TC) is used for the bottom electrodes. The bottom electrodes are in crossing direction to the top electrodes, forming a matrix for pixel-array addressing. The active semiconductor layer is sandwiched between the top and bottom electrodes. Detector pixels are separated by low refractive index dielectric materials (Liu *et al.* 2007). In our simulation, only the calculation of the photocurrent generated is considered, thus the electrodes have been removed in the simulation schematic.

In the simulation structure, the length of NPD pixels is set to be $3\mu m$ to investigate the optical power coupling between pixels, although in practical fabrication, the length of NPD pixels is only a few hundred of nanometers. The semiconductor fingers (pixels) play an important role in detecting incident field, which are converted into the photocurrents via the mechanism shown in Fig. 1.

The photocurrent generated in each NPD pixel can be quantitatively calculated via the following formula, which is directly derived based on the definition of current and mechanism shown in Fig. 1.

$$I_{ph} = \frac{q}{t_{sim}} N = \frac{q}{t_{sim}} \left( \sum_{pixel} N_2 \right) \cdot N_{density} \cdot A \cdot H, \tag{1}$$

where $q = 1.6 \times 10^{-19} C$, $t_{sim}$ is the total time used for simulation, $N$ is the total number of electrons, $N_2$ is the normalized number of electrons on the level $|2\rangle$ in a FDTD pixel, $N_{density}$ is the number of electrons per unit volume, $A$ is the area of the FDTD pixel, and $H$ is the height of the NPD pixels. Here, we set $t_{sim} = 1.0 ps$, $N_{density} = 0.563 \times 10^{22} / m^3$, $A = dx \times dy = 5nm \times 5nm = 25nm^2$, and $H = 300nm$, which is used in real NPD fabrications.

## 3. Near-field imaging by NPD array

In Fig. 2, we show a typical NPD array geometry, where the center-to-center distance between the NPD pixels is *w+s* with an inter-pixel gap of *s*. For an exemplary simulation to be shown below, we will show the case of *s*=60nm and *w*=90nm. The operating incident wavelength is 1550nm. The spatial resolution of the NPD is defined by the full-width half-maximum (FWHM) of the spatial distribution of the photocurrent response when it is illuminated by a near-point source. To generate the near-point source, we use a metal sheet with a small aperture having a small width (*sw*). The source is placed at a certain distance *sd* away from the front side of the center pixel of the NPD array. The optical absorption is calibrated to be 0.5/$\mu$m for a typical III-V semiconductor material, which corresponds to $N_{density} = 0.563 \times 10^{22} / m^3$ in eq. (1). In order to investigate the optical power coupling between NPD pixels, the average optical power in each pixel is calculated. In this initial 2D simulation, we assume a detector slab that is infinite in the direction perpendicular to the paper and the incident source has electric field polarization pointing along this infinite direction (we call this TM field). Fig. 3(a) shows the normalized field pattern, which indicates electric field quasi-guided by the center pixel (pixel 0) with subsequent coupling to the adjacent pixels (pixel 1, 2) and then to the next adjacent pixels (pixel 3, 4). Fig. 3(b) shows the corresponding photoelectron density from the electric field pattern of Fig. 3(a) with an arbitrary normalized linear scale. Fig. 4 shows the photocurrents in each pixel from the spatial distribution of photoelectron density profile of Fig. 3(b) using eq. (1). The estimated spatial resolution for this particular NPD array geometry is about 150nm, which corresponds to a resolution of $\lambda$/10.

## 4. Conclusion and future work

We investigated a new simulation method for photodetection in semiconductor medium with its application to a subwavelength resolvable NPD array, where a MLME-FDTD model was employed for the simulation. The FDTD simulations show us the optical power coupling between the NPD pixels, the spatial distributions of the electric field, and the photoelectron density of the proposed photodetector structure, from which the photocurrents can be calculated. This novel type of photodetector shows a high optical imaging resolution that is substantially below the diffraction limit, which can be potentially applied to the observation of nanoscale moving objects or living cells. Prototypes of such novel NPD devices have been successfully developed and characterized by us (Liu *et al.* 2007). Further parameter study such as width of the metal slit, polarization of incident light, distance between slit and the NPD array, etc will be conducted and reported soon.

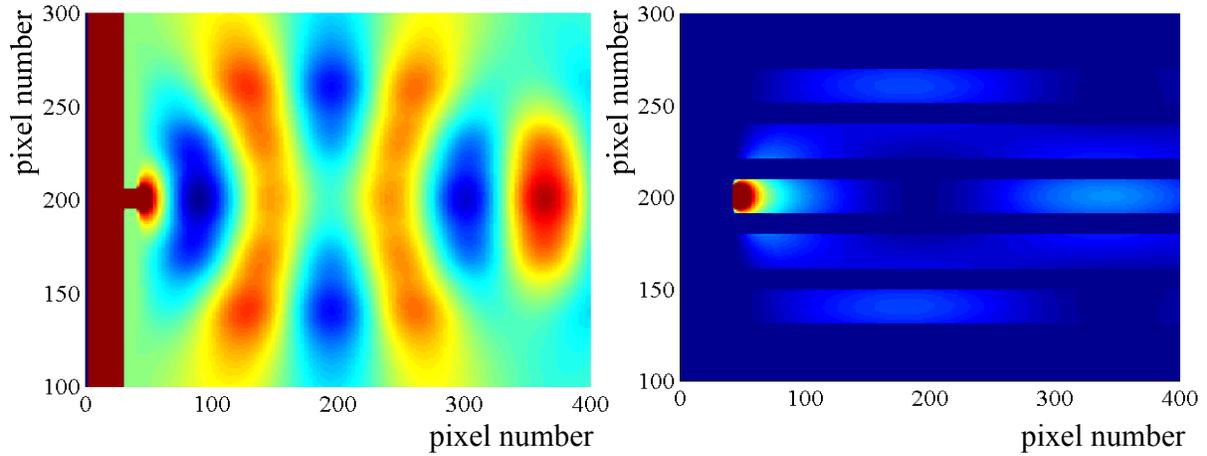

Fig. 3. Electric field pattern (left) and corresponding normalized photoelectron density (right). Ticks on axis represent the number of FDTD pixels (*dx* and *dy*), which are equal to 5nm in our simulation. The red color indicates higher amplitude in arbitrary linear scale.

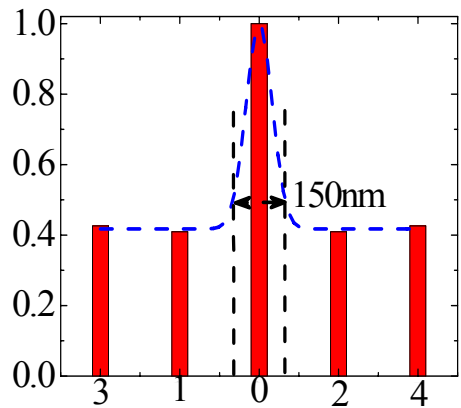

Fig. 4. Photocurrents in each NPD pixel. The effective FWHM spatial resolution is about 150nm, which corresponds to a resolution of $\lambda/10$.

**Acknowledgements**

This work was supported by NSF under Award No. ECS-0501589 and ECCS 0622185, by NSF MRSEC program under grant DMR-0076097, by the National Center for Learning & Teaching in Nanoscale Science and Engineering (NCLT) under the NSF Grant No. 0426328, and by the NASA Institute for Nanoelectronics and Computing under Award No. NCC 2-1363. This work was also supported by the Korea Research Foundation Grant funded by the Korean Government (MOEHRD). (KRF-2006-214-D00064).